\documentclass{JHEP3}

\newcommand{\be}{\begin{equation}}
\newcommand{\ee}{\end{equation}}
\newcommand{\bea}{\begin{eqnarray}}
\newcommand{\eea}{\end{eqnarray}}
\newcommand{\eref}[1]{(\ref{#1})}
\renewcommand{\(}{\left (}
\renewcommand{\)}{\right )}

\newcommand{\e}{{\rm e}}
\newcommand{\Or}{\mathcal{O}}
\renewcommand{\Re}{\mathop{\rm Re}}

\newcommand{\mpl}{m_{\rm Pl}}
\newcommand{\gs}{\delta_{\scriptscriptstyle \rm GS}}
\newcommand{\eps}{\epsilon}

\title{Warping and F-term uplifting}

\author{Philippe Brax \\
Service de Physique Th\'eorique, CEA/DSM/SPhT,
Unit\'e de recherche associ\'ee au CNRS, CEA--Saclay F--91191
Gif/Yvette cedex, France\\
E-mail: \email{brax@spht.saclay.cea.fr}}

\author{Anne-Christine Davis \\
Department of Applied Mathematics and Theoretical Physics,
University of Cambridge,
Clarkson Road, UK \\
E-mail: \email{A.C.Davis@damtp.cam.ac.uk}}

\author{Stephen C.~Davis, Rachel Jeannerot \\
Instituut-Lorentz for Theoretical Physics, Postbus 9506,
2300 RA Leiden, The Netherlands \\
E-mail: \email{sdavis@lorentz.leidenuniv.nl} ,
\email{jeannero@lorentz.leidenuniv.nl}}

\author{Marieke Postma \\
Nikhef, Kruislaan 409, 1098 SJ Amsterdam, The
Netherlands \\
E-mail: \email{mpostma@nikhef.nl}}

\abstract{We analyse the effective supergravity model of a warped
compactification with matter on D3 and D7-branes. We find that the
main effect of the warp factor is to modify the $F$-terms while
leaving the $D$-terms invariant. Hence warped models with moduli
stabilisation and a small positive cosmological constant resulting
from a large warping can only be achieved with an almost vanishing
$D$-term and a $F$-term uplifting. By studying string-motivated
examples with gaugino condensation on magnetised D7-branes, we find
that even with a vanishing $D$-term, it is difficult to achieve a
Minkowski minimum for reasonable parameter choices.
When coupled to an ISS sector the AdS vacua is uplifted, resulting
in a small gravitino mass for a warp factor of order $10^{-5}$.}


\preprint{arXiv:0707.4583}

\begin{document}

\section{Introduction}

If supersymmetry is realised in nature, it must be broken. Low scale
supersymmetry (SUSY) breaking is phenomenologically favoured: it addresses the
hierarchy problem, and leads to gauge coupling unification and
radiative electroweak symmetry breaking. If indeed the scale of SUSY
breaking is low, and the LHC sees the superpartners of the standard
model fields and measures their masses, what would this tell us about
the underlying microphysics?  String theory has tried to answer this
question since its early days~\cite{din}, but only recently has real
progress been made thanks to a better understanding of the moduli
sector and the associated sources of SUSY breaking.

Cosmological observations indicate that our universe undergoes a phase
of accelerated expansion. This motivates the search for de Sitter
vacua of string theory.  KKLT have found an explicit way of
constructing de Sitter solutions with a small cosmological constant in
the context of type IIB string theory, using a combination of fluxes,
D-branes and non-perturbative effects~\cite{KKLT}. In their proposal,
background RR and NS fluxes stabilise the complex structure moduli of
a Calabi-Yau compactification at some high scale. The remaining
K\"ahler moduli
can then be described by some low energy effective
theory.  Non-perturbative effects are invoked to stabilise the
K\"ahler moduli.  The resulting vacuum is SUSY preserving and anti de
Sitter.  An anti-D3-brane is added to uplift this solution to a dS
vacuum with broken SUSY.

Since the original KKLT paper many variations to their stabilisation
mechanism have been put forward.
Instead of an $\overline{\rm D3}$ brane, a
$D$-term originating from an anomalous $\mathop{\rm {}U}(1)$ could be
used for uplifting~\cite{BKQ}. This idea was put in
a manifestly gauge invariant form in~\cite{ana}, which
circumvents the problems of using $D$-terms for
uplifting~\cite{nilles,alwis}. The great advantage of an uplifting
$D$-term over the KKLT proposal is that the whole setup is
supersymmetric, with SUSY broken spontaneously in the vacuum; in
contrast, the $\overline{\rm D3}$ brane employed in KKLT breaks SUSY
explicitly. But the price to pay is that the uplifting term is
naturally of order one in Planck units, giving rise to a large
gravitino mass $m_{3/2} \sim \mpl/(8\pi)$ and thus to high scale SUSY
breaking. Uplifting mechanisms using $F$-terms have also been
analysed~\cite{ISS,dm,lebedev}, and models exist that are
supersymmetric and have a small gravitino mass~\cite{Okklt,lalak}.

In this paper we consider these issues within the context of warped
compactifications~\cite{gkp,quevedo,giddings} and study the effects of the
warping on the K\"ahler potential. A central assumption in this work
is that the throat dominates the volume of the 6D compactified
space. We find that the $F$-terms are warped down while the $D$-terms
are warping independent. Hence moduli stabilisation models with
consistent $D$-terms and a small positive cosmological constant
resulting from a large warping are only obtainable for vanishingly
small $D$-terms. The resulting models have an $F$-term uplifting. We
apply this setup to the case of string inspired configurations with
D7-branes.

The requirement of gauge invariance constrains the possible setup
considerably. In a general compactification, the non-perturbative
effects needed to stabilise the volume modulus can come from
either gaugino condensation on D7-branes or from D3 Euclidean
instantons.  The stack of D7s on which gaugino condensation occurs
is to be placed in the bulk. It cannot be in the throat, as then
the gauge coupling of the Yang-Mills theory living on the D7s is
red-shifted, leading to a vacuum with an exponentially small
modulus VEV, and the effective field theory breaks down. Gauge
invariance of both the superpotential from gaugino condensation
and the $D$-term from fluxes is only possible if there are charged
chiral fields living on the branes. This is achieved by
considering magnetised D7-branes. In addition to the moduli
fields, we should consider the origin of the standard model
fields. The standard model quark and lepton (super)fields, which
are chiral and in bi-fundamental representations, correspond to
strings stretching between stacks of D7-branes
\cite{intersecting}.  Gauge couplings of order one dictate that
the standard model branes are located in the bulk region. As
usual, the supersymmetry breaking in the moduli sector is
transferred to the superpartners of the standard model fields
through gravitational interactions.

We study the string inspired examples where the moduli stabilisation
superpotential results from the presence of magnetised D7-branes with
a chiral spectrum and gaugino condensation~\cite{dm}.  We find no
consistent Minkowski solutions with vanishingly small $D$-terms, as
needed for low scale SUSY breaking. However, starting from a stable
AdS vacuum with $D\approx 0$ (for example the SUSY AdS vacuum with
$F=D=0$), we find that an ISS uplifting can be applied leading a to small
gravitino mass for a reasonable warp factor.

This paper is organised as follows. In the next section we perform the
dimensional reduction of a warped space-time from ten to four
dimensions, and derive the form of the K\"ahler potential. In
addition, we discuss how the various elements in the effective action
--- the K\"ahler potential, the superpotential, the $D$-term, and the
gauge couplings --- red shift due to the warping.  In
section~\ref{s:moduli} we study moduli stabilisation with consistent
$D$-term and $F$-term uplifting.  In section~\ref{s:lift} we apply
these results to string inspired $F$-term uplifting models. The case
of an ISS uplifting is considered in this section.
We end with some concluding remarks.

\section{Dimensional reduction}
\label{s:dimred}

In this section we discuss dimensional reduction in warped
compactifications. This allows us to derive the form of the various
terms in the low energy effective four-dimensional theory, and in
particular how these terms are affected by the warping.

\subsection{The K\"ahler potential}
\label{s:Kahler}

The background space-time metric arising in warped IIB theories
preserving $N=1$ SUSY is of the form~\cite{gkp}
\begin{equation}
ds^2_{10} = \frac{\alpha'}{\e^{2A(y)}R^2(x)} g_{\mu\nu} dx^\mu dx^\nu
+ R^2(x) \e^{2A(y)} g_{m n} dy^m dy^n
\label{g10}
\end{equation}
with the length $x^\mu$ parameterising our four-dimensional world, and
the dimensionless $y^m$ the 6 extra dimensions. We scale $y^m$ so that
$V_6 \equiv \int d^6 y\sqrt{g_6} = (2\pi)^6$, where $g_6=\det g_{mn}$.
The radius $R(x)$ represents the only K\"ahler modulus and measures
the size of the extra dimensions.

The warp factor $\e^{A(y)}$ is present due to the non-zero fluxes on
the compactification manifold. The manifold is assumed to possess two
different regions. First of all, in the bulk the metric $g_{mn}$ and
the warp factor are of order one. In the throat, the metric is shifted
in such a way that $\e^{A(y)}$ is exponentially large. Throughout this work we
assume that the throat dominates the volume of the 6D compactified
space. We model the throat as a region akin to the AdS bulk of the
Randall-Sundrum model with one preferred direction. Along the throat,
the warp factor is essentially independent of
the transverse coordinates to the throat direction. In the tip of the
throat solution constructed by Giddings, Kachru and Polchinski,
$\e^{A_m} \approx \e^{2\pi K/3M g_s}$ with M units of R-R flux and K
units of NS-NS flux~\cite{gkp}. $g_s$ is the string coupling constant.
We also assume that the throat direction realises a
foliation of the manifold in such a way that compact cycles in the
transverse directions to the throat exist. In particular, we will
assume that one can wrap D$p$-branes around $(p-3)$ cycles located at an
essentially given value of $y$ along the throat, we will describe
these branes as being localised in the throat.  As we will ultimately
require the D7-branes to be located in the bulk, this assumption is
not essential.

Using the above metric~\eref{g10}, we obtain the effective
four-dimensional theory by integrating over the extra dimensions.  The
10D Ricci curvature of the full space-time metric contains a term
proportional to the curvature $R_4$ of $g_{\mu\nu}$ of the form
$R_{10} \supset (R^2 \e^{2A}/\alpha') R_4$. The remaining terms are
total derivatives which can be dropped in the effective action.  This
implies that the dimensionally reduced Einstein-Hilbert action, in the
supergravity frame, is
\be
\frac{1}{(2\pi)^7 \alpha'^4 g_s^2} \int d^{10} x \sqrt{-g_{10}}\, R_{10}
\supset  
\frac{1}{2\kappa_0^2} \int d^{4}x \sqrt{-g_4} \Big( f(T,\bar T)
R_4 + 6\partial_{T}\partial_{\bar T} f \partial_\mu T \partial^\mu \bar T
\Big)\, ,
\label{EH} \ee
where $g_4=\det g_{\mu\nu}$. We have introduced the scale
$\kappa_0^2 =\alpha' g_s$, and the moduli field $T$, which is related
to the size of the compactified dimensions
\begin{equation}
\Re T = \frac{ R^4}{2\pi g_s \alpha'^2} \, .
\label{T}
\end{equation}
We also introduce  $A_0$, defined by $\int d^6y \sqrt{g_6}
\e^{4A} = V_6 \e^{4A_0}$. Assuming that the warp factor is monotonic along the
throat, there will then exist a point $y_0$ such that $A(y_0) = A_0$.
Two natural situations can be envisaged.
For bulk domination, $A_0\approx 0$ while for throat domination we
have $A_0\approx A_m$~\cite{klebanov}. We will focus on the latter case,
which is consistent for $T \ll \e^{4A_0}$~\cite{giddings}. We find
\begin{equation}
f(T,\bar T) = e^{4A_0}(T+\bar T) + \cdots
\label{f}
\end{equation}
to leading order in the gravitational constant.

It is useful to define the Einstein frame, where
there are no field dependent couplings in the Einstein-Hilbert
action. The K\"ahler potential there is given by
$K = -(3/\kappa_0^2) \ln f$. The 4D
Einstein-frame metric is related to the supergravity frame metric by
the conformal transformation
\be
g^E_{\mu\nu} =  f g_{\mu\nu}
= \frac{e^{4 A_0} R^4}{\pi g_s \alpha'^2}g_{\mu\nu}
\, ,
\label{weyl}
\ee
which brings the gravitational action~\eref{EH} in to the form
$(2\kappa_0^2)^{-1}\int d^4x \sqrt{-g_E} R_4^{(E)}$.
Although ultimately we will want to work in the Einstein frame,
for the time being, we will remain in the supergravity
frame to calculate the corrections to $f$.

\subsection{Coupling to matter}

Let us now introduce matter in this setting. This will induce
subleading terms in the K\"ahler potential.  Consider fields living on
a D$p$-brane wrapped around a $(p-3)$-cycle of the 6-manifold, with
$p=(3,5,7,9)$ in type IIB string theory. Define by $d=(9-p)/2$ the
complex dimension of the complex normal bundle to the $(p-3)$ cycle.
Split the coordinates as $(x^\mu,z^i,z^a)$, $\mu=0\dots 3,\ i=1\dots
(p-3)/2, \ a=1\dots d$. The D-brane action is obtained from the
Dirac-Born-Infeld action at leading order
\bea S_{{\rm D}p}\supset - \frac{1}{g_{p+1,{\rm YM}}^2} \Bigg(&& \int
d^{p+1}x \sqrt{-\tilde g_{p+1}} \, \tilde g_{p+1}^{\mu\nu} \tilde
g_{p+1}^{\rho\sigma} F_{\mu\rho}F_{\nu\sigma} \nonumber \\ && {}
-\int d^{p+1} x \sqrt{-\tilde g_{p+1}} \hat g^{p+1}_{a\bar
b}(\Phi^a,\bar \Phi^{\bar b}) \tilde g^{\mu\nu}\partial_\mu \Phi^a
\partial_\nu \bar \Phi ^{\bar b} \Bigg) \, .
\label{SDp1}
\eea
The gauge coupling constant for the Yang-Mills (YM) field living
on the brane is $g^{-2}_{p+1,{\rm YM}} = (2\pi \alpha')^2 \tau_p$
with $\tau_p$ the brane tension; for a BPS brane it is
$\tau_p^{-1} = g_s(2\pi)^p \alpha'^{(p+1)/2}$. The gauge field is
$A_\mu$ along the non-compact dimensions and the $\Phi^a$'s are
transverse modes to the brane parameterising the normal bundle,
hence the contravariant indices. These fields are promoted to be
$x$-dependent only. The induced metrics are $\tilde g_{p+1}$ on
the $(p-3)$-cycle and $\hat g_{p+1}$ on the normal bundle.
Effectively we have $\tilde g_{\mu\nu,p+1}= \alpha' R^{-2}e^{-2A}
g_{\mu\nu}$ and $\hat g_{a\bar b}^{p+1}= (R^2 e^{2A}/\alpha')
g_{a\bar b}^{p+1}$. The warp factor is evaluated on the brane and
$g_{a\bar b}^{p+1}$ is the metric $g_{mn}$ induced on the normal
bundle, with a K\"ahler potential $g_{a\bar b}^{p+1}=\partial_a
\partial_{\bar b} g^{p+1}$.

Dimensional reduction of \eref{SDp1} leads to a 4D Yang-Mills
action $g_{\rm YM}^{-2}\int d^4 x \sqrt{-g_4} F^2$ with
\begin{equation}
\frac{1}{g^2_{\rm YM}}
=\frac{ R^{(p-3)} \tilde V_{p-3}}{2\pi g_s \alpha'^{(p-3)/2}} \, ,
\end{equation}
where we have defined $\tilde V_{p-3}= (2\pi)^{3-p} \int d^{p-3} y
\sqrt{g_{p-3}} \e^{(p-3) A(y)}$, which is dimensionless in our
conventions. We approximate this as $\tilde V_{p-3} = k_{p}
\e^{(p-3)A_b}$, where $k_{p}=\Or(1)$ and $A_b$ depends on the position
of the brane in the compactified dimensions. We will focus on two
different situations. First of all, the brane may be in the bulk where
the warp factor is close to unity, so $A_b \approx 0$. Another
situation corresponds to branes in the throat, then $A_b =A_m \approx
A_0$.  Supersymmetry requires the gauge coupling to be the real part
of a holomorphic function $g_{\rm YM}^{-2}=\Re f_G$, implying only a
D3/D7 system is supersymmetric. For these two cases
\begin{equation}
f_G = \left\{ \begin{array}{cc}
{\displaystyle \frac{k_3}{2\pi g_s}} & \qquad (p=3) \vspace{0.1in} \\
k_7T \e^{4 A_b} &  \qquad (p=7) \end{array} \right.
\label{fG7}
\end{equation}
with $A_b = 0$ for bulk branes and $A_b=A_m$ for branes in the throat.
In the D7 case, the warp factor implies that the gauge coupling is
very small if the branes live in the throat. As a consequence the
branes carrying the standard model fields must live in the bulk.

Notice that, as we assume that both the complex structure moduli and
the dilaton have been fixed by the flux induced superpotential (see
next subsection), the gauge coupling function is field independent on
D3-branes implying that no gaugino masses can appear there.  Strictly
speaking, the dilaton dependence of the D3 gauge kinetic function
$f_G\propto S =(2\pi g_s)^{-1}$ can lead to gaugino masses when the
dilaton has a non-zero $F^S$ term, i.e., contributes to the
supersymmetry breaking. Here we assume that $F^S =0$ and supersymmetry
breaking occurs only after dilaton stabilisation leading to gaugino
masses on D7-branes. Another advantage of D7s is the possibility of
having intersecting branes. On these intersections, strings stretching
between branes are in bi-fundamental representations, which can be
matched with the standard model representations in an easier
fashion. We will come back to intersecting branes at the end of this
section.

Consider matter fields on the branes. Scalar fields on the brane are
sections of the normal bundle to the brane belonging to the adjoint
$\mathop{\rm {}U}(N)$ representation for a stacks of $N$ D$p$-branes
(multiple gauge groups are present at intersections of
D$p$-branes). Their action~\eref{SDp1} reduces to
\be \frac{(\pi g_s)^{(p-7)/4}}{2} \int d^4 x
\sqrt{-g_4} (T+\bar T)^{(p-3)/4}\bar k^{(p)}_{a\bar b}
\partial \Phi^a \partial \bar \Phi ^{\bar b} \, ,
\label{SDp2}
\ee
where
\begin{equation}
\bar k^{(p)}=  \int
\frac{d^{p-3}y}{(2\pi)^{p-3}} \sqrt {g_{p-3}}\, \e^{(p-3)(A-A_0)}
g^{p+1}(\Phi^a, \bar \Phi^{\bar b}) \approx \e^{(p-3) (A_b-A_0)}
k^{(p)}(\Phi^a, \bar \Phi^{\bar b}) \, ,
\end{equation}
with $\hat g^{p+1}_{a\bar b} = (R^2 e^{2A}/\alpha') \partial_a
\partial_{\bar b} g^{p+1}$ the K\"ahler metric on the normal bundle,
and
\begin{equation}
k^{(p)}(\Phi^a, \bar \Phi^{\bar b}) =
\int \frac{d^{p-3}y}{(2\pi)^{p-3}} \sqrt {g_{p-3}}\,
g^{p+1}(\Phi^a, \bar \Phi^{\bar b}) \, . \label{kmetric1}
\end{equation}
As before $A_b = 0$ for branes located in the bulk, and $A_b=A_m$ for
branes in the throat.  Including \eref{SDp2} in the full supergravity
frame action \eref{EH}, we find
\begin{equation}
f = e^{4A_0}(T+ \bar T) - \frac{\kappa_0^2}{6} (T+\bar T)
 \e^{4A_b} k^{(7)} (\Phi^a, \bar \Phi^{\bar b})
- \frac{\kappa_0^2}{2\pi g_s} k^{(3)}(\Phi^a, \bar \Phi^{\bar b}) \, .
 \label{f2}
\end{equation}

Finally, consider matter fields living at the intersections of D$p$-
with D$p'$-branes. These matter fields live on a submanifold of
dimension $p+p'-12$. They are charged under the gauge groups of both
stack of branes, and therefore belong to bi-fundamental
representations. Their contribution to the K\"ahler potential can be
deduced from the previous result~\eref{SDp2}, with the change $p-3 \to
p+p'-12$. On the intersection, the induced metric reads $ \tilde
g^{p\cap p'}_{a\bar b}= (e^{2A}R^2/\alpha') g^{p\cap p'}_{a \bar b}$,
and $g^{p\cap p'}_{a\bar b}= \partial_a\partial_{\bar b} g^{p\cap p'}$
is the K\"ahler metric on the normal bundle to the brane
intersection. For intersecting D7-branes $p=p'=7$, located at $y=y_s$
we obtain
\be f = e^{4A_0}(T+\bar T) - \frac{\kappa_0^2}{6
(\pi g_s)^{1/2}} (T+\bar T)^{1/2} \e^{2A_{sm}}k^{(7\cap 7)}(\Phi^a,
\bar \Phi^{\bar b})
 \label{f3}
\ee
where $A_{sm} = A(y_{sm})$, and
\begin{equation}
\e^{2 A_{sm}} k^{(7 \cap 7)} (\phi^a, \bar \phi^{\bar b}) =
\int d^{2}y \sqrt {g_{2}} \e^{2A}
g^{7\cap 7}(\Phi^a, \bar \Phi^{\bar b}) \, .\label{kmetric2}
\end{equation}

Having derived the expression for $f$, the Einstein frame K\"ahler
potential is $K=-(3/\kappa_0^2) \ln (f)$.  For a
setup with a single D7-brane at $y=y_b$, and two intersecting D7-branes
at $y=y_{sm}$, the K\"ahler potential obtained from \eref{f2} and
\eref{f3} is
\bea K= - \frac{12 A_0}{\kappa_0^2}-
\frac{3}{\kappa_0^2} \ln & \Biggr[&
(T+\bar T )\bigg( 1 -\frac{\kappa_0^2}{6}
\e^{4(A_b-A_0)} k^{(7)} (\Phi^a, \bar \Phi^{\bar b}) \nonumber \\
&& {} - \frac{\kappa_0^2}{6 (\pi g_s)^{1/2}} (T+\bar T)^{-1/2}
\e^{2(A_{sm}-2A_0)}k^{(7\cap 7)}(\Phi^a, \bar \Phi^{\bar b}) \bigg)\Biggr] \, .
\label{K1}
\eea
This is the main result of this section. In terms of modular weights
in the heterotic notations we find
\begin{equation}
n_{\rm D7}=0,\quad n_{\rm D3}=-1, \quad n_{{\rm D7} \cap {\rm
D7}}=-\frac{1}{2}
\end{equation}
in agreement with the results of~\cite{ibanez}. The weights have a
geometrical origin.

\subsection{Fayet-Iliopoulos term}

Our model, to be discussed in section~\ref{s:moduli}, possesses an
anomalous $\mathop{\rm {}U}(1)$ symmetry. Stabilising the K\"ahler moduli is
achieved in the presence of a field-dependent Fayet-Iliopoulos.
Here we briefly recall the microscopic origin of such a
term~\cite{dsw,FI,quevedoT}, focusing on how it is affected by warping. The
existence of chiral fields leading to non-perturbative
superpotentials on the branes is linked to a  non-vanishing gauge
field along the magnetised brane compactified directions. Hence in
this context,  potential terms due to the magnetic fields are always
present. We will see that these terms arise in the 4D description
due an anomalous $\mathop{\rm {}U}(1)_X$ symmetry.

 It is most convenient to work
in the Einstein frame in order to obtain the Einstein frame
potential induced by the brane gauge fields. Ignoring all fields
except $T$, the 10D uplift of the Einstein frame metric is
\begin{equation}
ds^2 = \e^{-2 A(y) -4A_0} \frac{\pi g_s
{\alpha'}^3}{R^6(x)}g_{\mu\nu}^E dx^\mu dx^\nu + \e^{2A(y)}
R^2(x)g_{nm}dy^n dy^m
\end{equation}
which upon dimensional reduction gives the Einstein-Hilbert action
$(2\kappa_0^2)^{-1}\int d^4x \sqrt{-g_4} R_4$.  The magnetic flux
contribution to the action reads
\begin{equation}
S_{\rm YM} = -\frac{1}{g^2_{{\rm YM},8}} \int d^8x \sqrt{-\tilde
g_8} \tilde g^{nl}\tilde g^{mk}F_{nm}F_{lk} \, .
\end{equation}
Using $\sqrt{-\tilde g_8}\tilde g^{nl}\tilde g^{mk}
\propto R^{-12} \e^{-4A(z)-8A_0}$  gives, for a brane located
at $y = y_b$,
\be V_{\rm gauge} =
\frac{e^{-4A(y_b)-8A_0}}{2 \pi^2\kappa_0^{4} (T+\bar T)^3}
 \int d^4y \sqrt{g_4(y)}
F_{nm}F^{nm} \, .
 \ee
The potential can be identified with a $D$-term potential coming from an
anomalous $\mathop{\rm {}U}(1)$ gauge symmetry. Under this 
$\mathop{\rm {}U}(1)$ the volume modulus transforms 
$T \to T - i (\gs/2) \epsilon$ with $\epsilon$  infinitesimal gauge parameter,
giving rise to a $D$-term $V_D = (\gs^2 K_T)^2 /(8 {\Re f_G})$ with the gauge
kinetic function for a D7 $f_G$ in \eref{fG7} and $K_T = \partial_T K$ follows
from \eref{K1}. It follows that the Green-Schwarz parameter is
\be \gs = \frac{\sqrt{2 k_7}}{3 \pi \kappa_0^{2}} \e^{-4A_0} \(\int d^4y
\sqrt{g_4(y)} F_{nm}F^{nm}\)^{1/2} \label{gs} \, . \ee
The warping dependence of $\gs$ will also be obtained from anomaly
mediation conditions \eref{anom1}, \eref{anom2}. Consistency will then
fix the warping dependence of the above magnetic field integral.

\subsection{Superpotential}
\label{s:W}

The presence of fluxes induces a superpotential for the dilaton and
the complex structure moduli of the form~\cite{W}
\begin{equation}
W = \int G_3\wedge \Omega \, ,
\end{equation}
where the 3-form on the compactification manifold can be identified
with
\begin{equation}
\Omega_{nml} = \bar \eta \Gamma_{nml} \eta \, ,
\end{equation}
with $\eta$ a Killing spinor corresponding to the remaining
supersymmetry.  We have introduced $\{\gamma_n,\gamma_m\}= 2 g_{nm}$
and the warped Dirac matrices $\{\Gamma_n,\Gamma_m\} = 2G_{n m}$ with
$G_{nm}=e^{2A} g_{nm}$. The superpotential can be written as
\begin{eqnarray}
W = \int d^6y \sqrt{G_6} (\bar \eta \Gamma^{nml}\eta)(G_3)_{nml}
= \e^{3A_0} \tilde W_0 \, ,
\end{eqnarray}
where the constant is
\begin{equation}
 \tilde W_0 \approx \int d^6y \sqrt{g_6}  (\bar \eta \gamma^{nml}\eta)
(G_3)_{nml} \, .
\end{equation}
Here and in the following we use the notation that the tilde
quantities are independent of the warp factor, whereas the equivalent
quantity without a tilde has some warp factor absorbed in it.  To get
the above equation we used that the $\gamma$-matrices are of order
one, and in the second line that the integral is dominated by the
value of the integrand in the throat.

Let us now include matter on D-branes. The full superpotential is
\begin{equation}
W = \e^{3A_0} \tilde W_0 + \tilde W_{\rm SM}(\Phi^a) +
\tilde W_{\rm NP}(T,\Phi^i) \, .
\label{W1}
\end{equation}
The moduli stabilisation potential $\tilde W_{\rm NP}$, which involves
the volume modulus $T$ and chiral fields $\Phi^a$ charged under
the condensing gauge group, will be discussed in the next section.
The standard model lives on intersecting D7s in the bulk, and has
a superpotential
\begin{equation}
\tilde W_{\rm SM}= \frac{1}{2} \e^{A_{\rm sm}} \tilde
\mu_{ab}\Phi^a\Phi^b +\frac{1}{3}  \tilde \lambda_{abc} \Phi^a
\Phi^b \Phi^c \label{mu1}
\end{equation}
which contains the Yukawa couplings and a $\mu$-term. We perform a
K\"ahler transformation
\be K \to K + \frac{12 A_0}{\kappa_0^2}, \qquad W \to \e^{-6A_0} W
\label{Ktrafo}
\ee
which leaves the $N=1$ supergravity Lagrangian invariant.  This
removes the constant from the K\"ahler potential \eref{K1}.  Consider
further the small field approximation for the standard model fields
$k^{(7\cap 7)}_{a\bar b} = \delta_{a\bar b}$, and introduce the
rescaled fields $\phi^a = \Phi^a \e^{-2A_0} (4\pi g_s)^{1/4}$. This
gives a K\"ahler potential
\be K = -\frac{3}{\kappa_0^2} \ln  \Biggr[ (T+\bar T )\bigg( 1 -
\frac{\kappa_0^2}{3} (T+\bar T)^{-1/2} \sum_a |\phi^a|^2-
\frac{\kappa_0^2}{6} \e^{4(A_b-A_0)} k^{(7)} (\Phi^i, \bar
\Phi^{\bar j})
  \bigg) \Biggr] \, ,
\ee
where we have set $A_{sm} = 0$ appropriate for SM fields in the bulk.
The superpotential is
\bea
W &=& \e^{-3A_0} \tilde W_0 + \e^{-6A_0} \bigg( \tilde W_{\rm SM}(\phi^a) +
\tilde W_{\rm NP}(T,\Phi^i) \bigg) \nonumber \\
 &\equiv& W_0 +  W_{\rm SM}(\phi^a) +W_{\rm NP}(T,\Phi^i) \, .
\eea
As before, we use the notation that the tilde quantities are
independent of the warp factor, whereas the equivalent quantity without
a tilde has some warp factor absorbed in it.  Now $W_{\rm SM}=
\frac{1}{2} \mu_{ab} \phi^a \phi^b +\lambda_{abc} \phi^a \phi^b
\phi^c$ with
\begin{equation}
\mu_{ab}=\e^{-2A_0}(4\pi g_s)^{1/2} \tilde \mu_{ab} \, ,
\qquad
\lambda_{abc}=(4\pi g_s)^{3/4} \tilde \lambda_{abc} \, .
\end{equation}
The effective Yukawa couplings $\lambda_{abc}$, and thus the
standard model masses, do not depend on the warp factor $A_0$ explicitly.
The effective $\mu$ term red-shifts as $\e^{-2A_0}$. As this is too large
compared to the gravitino mass, we can instead use the
Giudice-Masiero mechanism to generate a $\mu$ term~\cite{giudice}.
Indeed, the gravitino mass $m_{3/2} = \e^{\kappa_0^2 K} \kappa_0^3
W \propto \e^{-3A_0}W_0$. This opens up the possibility to tune
the scale of supersymmetry breaking by ``switching on'' the
warping.

\subsection{The supergravity approximation}
\label{ss:sugra}

In this subsection we will discuss mass scales, and the regime where
the low energy effective supergravity action is to be trusted.

So far we have expressed all quantities in terms of the fundamental
string scale $\kappa_0 = \sqrt{\alpha' g_s}$. We can switch scales and
go to Planck units by simply changing by $\kappa_0 \to \kappa_4$ in all
formulae. For an observer located at $y=y_{sm}$, which we define to be
the location of the standard model fields, the 4D Planck mass is
\be
\mpl^2 \equiv \frac{1}{\kappa_4^2}=
\frac{f(T_0)}{\kappa_0^2} \frac{e^{2A(y_{sm})} R(T_0)^2}{\alpha'}
= \frac{1}{\alpha'} \(\frac{8 \pi T_0^3}{g_s}\)^{1/2} \e^{4A_{0}+2A(y_{sm})}
\, ,
\label{mpl}
\ee
where $T_0$ is the vacuum expectation value of $T$.  Note that this
change of length scale implies a conformal scaling of the spacetime
metric~\eref{weyl}, $g^E_{\mu\nu} \to \kappa_0^2/\kappa_4^2 \,
g^E_{\mu\nu}$, as well as the K\"ahler metrics~\eref{kmetric1} and
\eref{kmetric2}, $k^{(p)} \to \kappa_0^2/\kappa_4^2 \, k^{(p)}$.  For the
remainder of this paper we will make the replacement $\kappa_0 \to
\kappa_4$, and work in units of $\kappa_4$.  The K\"ahler potential
becomes
\be
K = -\frac{3}{\kappa_4^2} \ln \Biggr[ (T+\bar T )\bigg( 1
- \frac{\kappa_4^2}{3} (T+\bar T)^{-1/2} \sum_a |\phi^a|^2-
\frac{\kappa_4^2}{6} \e^{4(A_b-A_0)} k^{(7)} (\Phi^i, \bar \Phi^{\bar
j}) \bigg) \Biggr] \, .
\label{K2}
\ee

The supergravity approximation is valid when the Kaluza-Klein masses
of the higher dimension model are large and the higher order terms in
the supergravity Lagrangian are under control. The lightest KK
particles come from the tip of the throat where the length scales of
the extra dimensions are maximal.  Typically we expect that the KK
masses are red shifted
\begin{equation}
m_{KK}^{(n)} \approx  ne^{-2A_0} \mpl
\end{equation}
for $T=\Or(1)$~\cite{giddings}. Thus if $m$ is the typical mass scale in
the low energy effective theory, we require $m < e^{-2A_0} \mpl$.

Let us now turn to the higher order terms in the Lagrangian. We use
$R_{10}^n$ to denote $n$ contractions of the ten-dimensional
Riemann tensor symbolically (derivatives of the Riemann tensor can
be treated similarly). Typically, upon dimensional reduction in the
supergravity frame, these terms yield
\begin{equation}
R_{10}^n\approx \sum_{p=0}^n c_p e^{(4p-2n)A(y)} R_4^p R_6^{n-p}
\end{equation}
for some coefficients $c_p$. $R_6 \approx M_s^2$, where
$M_s^2 = 1/\alpha'$ is the string scale. Expanding
the ten-dimensional effective action gives
\bea
S_{\rm eff} &\approx& M_s^{10} \int d^{10} x\sqrt{-g_{10}}
\sum_{m=0}^\infty d_m \frac{R_{10}^m}{M_s^{2m}}
\nonumber \\
&\approx&  M_s^4 \int
d^4 x \sqrt{-g_4} \sum_{p=0}^\infty  c_p \frac{R_4^p}{M_s^{2p}}
\(\sum_{m=p}^\infty d_m \int d^6 y \sqrt{g_6}  e^{(4p-2m +2)A(y)}\)
\eea
with $d_0=0$. The sum over $m$ is dominated by the term $m=p$ and
scales like $\e^{(2p+2)A_0}$. Note that we retrieve the $\e^{4A_0}$
behaviour when $p=1$. Expanding the action, and switching to Planck
units gives
\begin{equation}
S_{\rm eff} \approx \mpl^4 \int d^4 x \sqrt{-g_4}
\frac{ R_4}{\mpl^2}\(1+ \e^{2A_0}\frac{ R_4}{\mpl^2} + \cdots \) \, .
\end{equation}
Hence the effective action in the supergravity frame after dimensional
reduction is a series expansion where higher order terms are
suppressed for $m^2 \e^{2A_0} \ll 1$, with $m$ the typical low energy
scale.  This is automatic in the regime where throat KK particles are
heavy. Thus, in general, we can neglect the higher order corrections
and work in the lowest order of the supergravity
approximation. Although $\alpha'$ corrections could help
with moduli stabilisation and uplifting, we do not expect them to be
useful for large warping. This is the case for the example in
Appendix~\ref{a:alpha}.

\vspace{0.25in}

To summarise, we have derived how the various terms in the low
energy effective four-dimensional theory are affected by the
warping. The K\"ahler potential picks up a constant piece
proportional to $A_0$, which originates from the warping
dependence of the 6D volume. This translates into a warping down
of the $F$-terms. On the other hand as we will find in the next
section, the $D$-terms are scaled down by the volume of the
4-cycle, which for bulk branes is warping independent. When trying
to find moduli stabilised configurations with a small cosmological
constant obtained thanks to a large warp factor, one cannot
accommodate large $D$-terms. The $D$-terms must be almost
vanishing. The resulting configurations are then $F$-term uplifted
vacua. In the next section, we will present string inspired
configurations with an $F$-term uplifted minimum; however, we find
that lifting the minimum all the way to Minkowski space proves
to be difficult.

\section{$F$-term uplifting with consistent $D$-terms}
\label{s:moduli}

So far we have considered a stack of branes with gauge group
$\mathop{\rm {}U}(N)$.  By separating the branes, i.e., giving VEVs to
the scalars in the adjoint representation, one can break the gauge
group to a product of $\mathop{\rm {}U}(N_i)$'s and $\mathop{\rm
{}U}(1)$'s. The rank is unaffected by the breaking and the model is
non-chiral as the original fields are in the adjoint
representation. In more general situations such as orientifold
compactifications one can obtain chiral spectra (see~\cite{ana}). For
instance, for a single brane with gauge group $\mathop{\rm {}U}(1)$, a
chiral field arises as the open string joining the brane and its
orientifold image. By taking a stack of $(N+1)$ branes and separating
one of the branes from the other ones, one can envisage a gauge group
$\mathop{\rm SU}(N) \times \mathop{\rm {}U}(1)$ with a chiral
spectrum. Quarks in the fundamental representation of $\mathop{\rm
SU}(N)$ arise as open strings joining the stack of $N$ branes to the
single $\mathop{\rm {}U}(1)$ branes, antiquarks appear too as open
strings joining the stack of branes to the orientifold image of the
$\mathop{\rm {}U}(1)$ brane. There is also a charged field coming from
the open string joining the $\mathop{\rm {}U}(1)$ brane to its
orientifold image~\cite{dm}.

The presence of chiral matter is intimately linked to the existence of
an internal flux on the $\mathop{\rm {}U}(1)_X$ brane, i.e., the brane
is magnetised. In the following we will consider a supergravity model
with such a matter content. In this case, with quarks and antiquarks
in the fundamental and anti-fundamental representations,
non-perturbative phenomena can occur leading to superpotentials
involving composite meson fields charged under the
$\mathop{\rm {}U}(1)_X$~\cite{din,gauginocondens}.  To be concrete, we take $N$
D7-branes, possessing a chiral spectrum of $N_f$ quark pairs
$\{\Phi_i, \tilde \Phi_i\}$ with charges $q_i$ and $\bar
q_i$~\cite{ana,quevedo}. In addition there is one $\mathop{\rm SU}(N)$
singlet $\Xi$ whose $\mathop{\rm {}U}(1)_X$ charge we normalise to
$-1$. We take the quark (and antiquark) charges to be equal, so $q_i =
q_1$ (and $\bar q_i = \bar q_1$).  The $\mathop{\rm {}U}(1)_X$ is
anomalous, the implications of which we will discuss in
subsection~\ref{s:D}.

\subsection{Moduli superpotential}
\label{ssec:W}

The effective, non-perturbative superpotential generated by the
gaugino condensation on the stacks of D7-branes is
\be \tilde W_{\rm NP} = \kappa_4^{-3} (N-N_f) \left(
\frac{e^{-8\pi^2 f_G}}{\kappa_4^2 \, {\rm det} (\Phi_i \tilde
\Phi^j)} \right)^{1/(N-N_f)} \ee
where $f_G$ is given by \eref{fG7} for $p=7$. The gaugino condensation
branes are located in the bulk with $A(y_b)=A_b$. Together with
$W_0$, this provides a stabilising potential for the volume
modulus $T$. In addition we introduce a direct coupling between
the chiral matter fields~\cite{dm}
\be \tilde W_{\rm m} = \kappa_4^{-1+q} \tilde m \ {\rm
det} (\Phi_i \tilde \Phi^j)^{1/N_f}\Xi^{q} \ee
where $\tilde m$ is constant and $q=q_1+\bar q_1$.
For simplicity we assume an overall squark condensate $\chi$, for
which $|\Phi_i|^2 = |\bar \Phi_i|^2 = |\chi|^2 \e^{4(A_0-A_b)}/N_f$. The
composite field $\chi$ then has charge $q/2$.
In the small field approximation
$k^{(7)} = \sum_i (|\Phi_i|^2 + |\tilde \Phi_i|^2) + |\Xi|^2$, where
we have defined $|\Xi|^2 = 2 |\zeta|^2 \e^{4(A_0-A_b)}$. The K\"ahler
potential then reduces to
\be K = -\frac{3}{\kappa_4^2} \ln  \Biggr[ (T+\bar T )\bigg( 1 -
\frac{\kappa_4^2}{3} (T+\bar T)^{-1/2} \sum_a |\phi^a|^2-
\frac{\kappa_4^2}{3} \(|\chi|^2 + |\zeta|^2\)
  \bigg) \Biggr] \, .
\label{K3}
\ee
The superpotential relevant for moduli stabilisation is
\bea
W &=& \e^{-3A_0} \tilde W_0 + \e^{-6A_0} \left[
\tilde W_{\rm NP}(T,\chi) + \tilde W_{\rm m}(\chi, \zeta) \right]
\nonumber \\
&\equiv& W_0 + W_{\rm NP}(T,\chi) + W_{\rm m}(\chi, \zeta)
\label{W2}
\eea
with
\bea
&&
W_{\rm NP} = A \frac{\e^{-a T}}{ \chi^{b}}= \frac{
\e^{-2(3+b)A_0+2 b A_b} }{\kappa_4^{b+3}} \tilde A \frac{\e^{-a T}}{\chi^{b}}
\ , \nonumber \\ &&
W_{\rm m} = m \zeta^{q} \chi^2
= \frac{\e^{-2(1-q)A_0-2(2+q)A_b}}{\kappa_4^{1-q}}
 2^{q/2}  \tilde m \, \zeta^{q} \chi^2 \, ,
\label{W3}
\eea
and
\be \tilde A = (N-N_f) N_f^{b/2}   \, ,\qquad a = \frac{8
\pi^2 k_N \e^{4A_b}}{N-N_f} \, ,\qquad b = \frac{2 N_f}{N-N_f}
\, . \label{Aab}
\end{equation}
As a result of the warping, the scales $A$ and $m$ in the
non-perturbative superpotential are no longer at the Planck scale.

The $F$-term potential is
\be
V_F = \e^{\kappa_4^2 K}
\left( D_I W K^{I \bar J} \overline{D_J W}
-3 \kappa_4^2|W|^2\right)
\ee
with $D_I = \partial_I W + \kappa_4^2 K_I W$.

\subsection{Anomaly cancellation and $D$-term potential}
\label{s:D}

The model possesses an anomalous $\mathop{\rm {}U}(1)_X$ symmetry whose
anomaly can be cancelled by the Green-Schwarz mechanism if $T$
transforms non-trivially under $\mathop{\rm {}U}(1)_X$. Under a gauge
transformation with infinitesimal parameter $\epsilon$, $\delta T =
-i(\gs/2)\epsilon$, where $\gs$ is the Green-Schwarz parameter. The
corresponding transformations for the quarks and antiquarks living on
the magnetised brane, and the $\mathop{\rm SU}(N)$ singlet, are
respectively $\delta \Phi_{i} = iq_1 \epsilon \Phi_{i}$, $\delta
\tilde \Phi_i = i\bar q_1 \epsilon \tilde \Phi_{i}$, and $\delta \Xi =
-i\epsilon \Xi$.  The required cancellation condition for the
$\mathop{\rm SU}(N)^2 \times \mathop{\rm {}U}(1)_X$ anomaly is
\begin{equation}
k_N \e^{4A_b}\gs=\frac{N_f}{4\pi^2} (q_1+ \bar q_1) \, , \label{anom1}
\end{equation}
and the one for the $\mathop{\rm {}U}(1)_X^3$  anomaly is
\begin{equation}
k_X \e^{4A_b} \gs=\frac{1}{6\pi^2}  (N N_{f} (q_1^3+\bar q_1^3)-1) \, .
\label{anom2}
\end{equation}
The $\mathop{\rm {}U}(1)_X$ $D$-term potential follows from $V_D = (2 \Re
f_X)^{-1} (i \eta^I K_I )^2$, with $I = \{T, \chi, \zeta \}$ and
$\eta = \{- i \gs/2,iq/2 \chi,-i\zeta\}$. Here $\eta^I \epsilon$ are
the infinitesimal gauge transformations of the fields.
Hence $V_D \propto \gs^2/k_X \propto \e^{-12A_b}$.  For bulk D7s
$A_b=0$, and $V_D$ is independent of the warp factor. This is a
crucial result as it implies that, for $g_{\rm YM} \sim 1$,  $D$-terms
must vanish in order for the gravitino mass to be warped down.

Using \eref{K3} the $D$-term potential is then
\begin{equation}
V_D = \frac{9\gs^2}{8\kappa_4^2 (T+\bar T)^2\Re f_{X}}
\( 1 +\frac{(T+\bar T)}{3\gs Y}\left[q |\chi|^2
- 2 |\zeta|^2\right]\)^2
\label{VD}
\end{equation}
with $q=q_1+\bar q_1$ and
\be  \qquad Y= 1-\frac{\kappa_4^2}{3} |\chi|^{2}
-\frac{\kappa_4^2}{3}|\zeta|^2 \, .
\ee

\section{Uplifting procedures}
\label{s:lift}

As discussed in section~\ref{s:moduli}, we consider an $N=1$ SUGRA
model with the following data:
\bea
K & =& -\frac{3}{\kappa_4^2}
\ln \left[ (T+\bar T)\(1-\frac{\kappa_4^2}{3}\{|\chi|^2 +|\zeta|^2\}\)
\right] \, , \nonumber \\
W &=& W_0 + A \frac{\e^{-aT}}{\chi^b} + m \zeta^q \chi^2 \, ,
\label{model}
\eea
and the $D$-term potential is given by~\eref{VD}, with the gauge kinetic
function $f_X = k_X T$.  Gauge invariance relates the parameters
$bq=a\gs$. The warping dependence of the superpotential is given by
\eref{W2}, \eref{W3}.

In this section we will look for (metastable) vacua of this system,
which have zero cosmological constant and low scale SUSY breaking.
Notice first that the $D$-term is warping independent whereas the
$F$-terms depend on the warping. Guaranteeing a small supersymmetry
breaking scale can only be achieved for a small value of the $D$-term
contribution.  This holds true even in the absence of warping.

The $D$-term can be (partially) cancelled by a non-zero VEV for
$\zeta$.  This generates a mass term for the $\chi$-fields.  If the
mass is greater than the other mass scales in the problem the
$\chi$-fields can be integrated out.  The resulting effective theory
involving only $T$ and $\zeta$ has a SUSY preserving $F=D=0$ AdS
minimum.  An additional $F$-term sector is needed to break SUSY and lift
this minimum to Minkowski. We discuss this in more detail in
subsection~\ref{s:iss} where we add an Intriligator-Seiberg-Shih 
(ISS)~\cite{ISS} section to the model.

It is clear then that to get a SUSY breaking Minkowski vacuum in the
model \eref{model}, we must be in the regime where the $\chi$-fields
are light and cannot be integrated out.  This is the approach taken in
the next subsection.  We will perform a systematic expansion of the
potential in a small parameter $\eps \propto T \chi^2/\zeta^2$.
However, we do not find a consistent solution either analytically or
numerically, at least not for credible parameter choices.
This strongly suggests that the only way to obtain a Minkowski vacuum
with a small gravitino mass in our model with $D$-terms, is by adding an
extra $F$-term lifting section.

Readers only interested in a working model can go immediately to
subsection~\ref{s:iss}.

\subsection{Uplifting and stabilisation with light quarks}
\label{s:exp}

As discussed above, we are looking for a $D\approx 0$ minimum allowing for the
possibility of low scale SUSY breaking.  Furthermore, we assume that the
quarks are light and cannot be integrated out. Taking inspiration from
\cite{dm}, we will consider the limit $\chi^2 \ll \mpl^2$ and $T\gg
1$. Then the dominant contributions to the potential are $V_D$ and
\be
V_0= \e^{K} \(K^{T\bar T} |D_T W|^2 - 3|W|^2\) = \frac{a A\e^{-aT}}{2
T^2 Y^3}\( \frac{A a T \e^{-aT}}{3\chi^{2b}} +\frac{W}{\chi^{b}} \)
\ee \be V_1 = \e^{K} K^{\chi\bar \chi} |D_\chi W|^2 =
\frac{3-\chi^2}{24 T^3 Y^2} \(2m \zeta^q \chi -b \frac{A
\e^{-aT}}{\chi^{b+1}} +\frac{\chi}{Y}W\)^2 \, .
\ee
The $K^{\zeta\bar \zeta} |D_\zeta W|^2$ contribution turns out to be
subdominant. All warping dependence has been absorbed into the
parameters $A, W_0, m$, and we work in units with $\kappa_4^2=1$.  To
leading order in $\chi$, $V_1$ reduces to the global SUSY
potential. Perturbatively, one can find a minimum of the potential
around the solutions of $D=0$ and $\partial_\chi W=0$ (i.e.\
$F_\chi=0$ for a global SUSY theory). Thus the zeroth order
approximate solution is
\begin{equation}
\zeta_0^2= \frac{3\gs}{4T_0} \, , \qquad
\chi_0^{b+2}= \frac{bAe^{-aT_0}}{2m \zeta_0^q}
\label{chiral0}
\end{equation}
with $T_0$ unspecified at this stage.  The above solution~\eref{chiral0}
then implies $W_{\rm m}^{(0)} = (b/2) W_{\rm NP}^{(0)}= m \epsilon
\zeta_0^{2+q}/(q a T_0)$.

Stabilising the volume modulus $T$ is the hard part.  Write
$T=T_0(1+\epsilon T_1 +\epsilon^2 T_2 +\cdots)$, and similarly for
$\chi$ and $\zeta$, with $\eps$ a small expansion parameter.  If,
following~\cite{dm}, we expand in $\eps = \chi^2/\zeta^2$ we find a
runaway behaviour for at least one combination of the fields.  To avoid
this we expand instead in
\be
1 \gg \epsilon \equiv \frac{q a T_0 \chi_0^2}{\zeta_0^2}
\gg \frac{1}{\sqrt{a T_0}}  \, .
\label{expansion}
\ee
We expand around
the lowest order solution \eref{chiral0}.  The details of this
expansion can be found in Appendix~\ref{a:stephen}; here we only
discuss the main results.  The first order corrections to $\chi,\zeta$
are fixed by $V_D$ and $V_1$, which then leaves $V_0$ to determine the
correction to $T$. To leading order, the remaining terms in
the potential (arising from $V_0$) are
\be
V^{(1)}_{\rm min} - \frac{3m^2 \epsilon^3\zeta_0^{2q}}{8
(2+b) a^2 T_0^5} \left\{
(2+b+4a + b q)
+\frac{a T_0 W_0}{m \epsilon \zeta_0^q} (8+4b+4a + b q)\right\} T_1  +
\cdots
\label{eV0}
\ee
with
\be
V^{(1)}_{\rm min} = \frac{3m^2 \epsilon^2 \zeta_0^{2q}}{8 a^2 T_0^5}
\(1 + 2 \frac{a T_0 W_0}{m \epsilon\zeta_0^{q}} \) \, .
\label{Vmin}
\ee
In general, we see that the above
potential has runaway behaviour for $T_1$, indicating that $T_0$
and the solution~\eref{chiral0} is not close to a minimum. The exception to
this is
\be \frac{a T_0 W_0}{m \epsilon \zeta_0^q} = -
\frac{2+b+b q +4a}{4(2+b)+b q +4a}
\label{mincond}
\ee
in which case the $T_1$ terms in \eref{eV0} cancel, leaving open the
possibility that $T_1$ could be stabilised by higher order terms. The
second order results, given explicitly in Appendix~\ref{a:stephen}, allow
for this possibility.  Thus if our solution is to be an extremum the
above condition must be satisfied. Note that this implies $W_0 \sim m
\epsilon \zeta_0^q/T_0$, i.e.\  $W_0 \ll m$; this is natural in a
warped background, see \eref{W2}, \eref{W3}.

For a special solution satisfying \eref{mincond} the potential at
the minimum is $V^{(1)}_{\rm min}$~\eref{Vmin}. Picking $T_0$ such
that $V^{(1)}_{\rm min}$ vanishes, and we have a Minkowski vacuum,
requires
\be 4a  \approx 2(2+b) - b q \, .
\label{cond}
\ee
For the parameters \eref{Aab}, with $A_b=0$, this implies $N \approx
8\pi^2 k_N + N_f q/2$. Satisfying this is only possible for large
values of $N$ or small $k_N$. The value of $T_0$ is
\be
a T_0 \approx
\frac{b+2}{2}\ln \frac{m q}{(-2W_0)}
+  \ln \frac{A b}{2m} +a\ln\frac{4}{3 \gs} \, .
\label{aT1}
\ee
Since $(aT_0)$ depends only logarithmically on the parameters it will
never be large for moderate values of $a$, $b$, etc., and so
it is extremely hard to satisfy the condition \eref{expansion} for which
our expansion is valid.  Indeed, scanning through parameter space it
follows that in the regime of validity of the expansion
\eref{expansion}, the solution \eref{cond}, \eref{aT1} requires very
large gauge groups $N \sim 10^2 - 10^3$.  We reject this
possibility on the grounds that the presence of so many D7-branes
would back react on the geometry, and invalidate the supergravity
results derived in section~\ref{s:dimred}, which motivate this model.

To conclude, we do not find a Minkowski vacuum with $D \approx 0$ for
light quarks, at least not for credible parameter choices.  The
obstacle to finding a metastable minimum is the
stabilisation of the volume modulus $T$.  Since our analytics only cover
part of parameter space, we have also searched numerically
for $D \approx 0$ solutions, but with negative results as well.  These
conclusions are independent of the amount of warping.

\subsection{Uplifting with an ISS sector}
\label{s:iss}

In this section we consider the case of heavy quarks.  $D \approx 0$
implies $\zeta \neq 0$, which generates a mass term for the
$\chi$-fields.  If their mass is greater than the other mass scales in
the problem the $\chi$-fields can be integrated out. The resulting
effective potential has a SUSY preserving AdS minimum.  We include an
ISS sector for the necessary uplifting.

But before we discuss this model in detail let us say a few words
about higher order $\alpha'$ corrections to the K\"ahler potential.
As noted in~\cite{quevedoT}, if these corrections are large the SUSY
conditions $F = D =0$ cannot be satisfied, and supersymmetry is
broken, possibly at a low scale.  However, this does not work for our
setup for two reasons.  First, the resulting minimum in our model is
dS.  And secondly, the $\alpha'$ corrections are exponentially small
in the presence of warping, and do not play a r\^ole. More details can
be found in Appendix~\ref{a:alpha}.

We can integrate out the heavy quark fields by solving the $F_\chi=0$
constraint in the global supersymmetry limit, i.e. $\partial_\chi
W=0$, leading to
\begin{equation}
\chi^{b+2}=\frac{bAe^{-aT}}{2m\zeta^q}
\end{equation}
and
\be W = W_0 + \(2+b\)\frac{A}{2}\(\frac{2m}{b A}\)^{\frac{b}{b+2}}
\zeta^{\frac{qb}{b+2}}\e^{-\frac{2a}{b+2}T}
\label{Wads}
\ee
with $W_0 \sim \e^{-3A_0}$. This system has a SUSY AdS minimum with
$F_T = F_\zeta=D=0$: The $D= 0$ constraint fixes $\zeta$, whereas
$F_T$ fixes $T$, and $F_\zeta =0$ follows then automatically as a
consequence of gauge invariance.  The superpotential is dominated by
the constant piece $W_0$.  Extra input is needed to break SUSY and
lift the solution to Minkowski. For definiteness, we will include an
ISS sector~\cite{ISS} for this purpose. In the absence of warping, the
coupling to an ISS sector has been studied in \cite{pok}. Working at
the level of an effective field theory description, we do not attempt
to present a full string-theoretic construction of the appearance of
an ISS sector in warped compactifications. As in \cite{pok}, we only
extract the ingredients which are necessary at the supergravity
level. For more details about the possible string theory
constructions, see \cite{oo,uf,ka}.

Consider a new contribution to the superpotential
\begin{equation}
\tilde W_{\rm ISS}= \tilde h \, {\rm tr} (\bar q M q) -  \tilde h
\tilde \mu^2 {\rm tr} (M) \, .
\end{equation}
The warping dependence of the coupling constants is left undetermined
as it depends on the details of the possible string realisation of the
model. We will find constraints on this warping dependence in order to
obtain a low gravitino mass. If these constraints are not satisfied in
explicit models of compactification, the resulting low energy
supergravity model with ISS uplifting will not lead to a LHC testable
phenomenology.  

The quark fields $q^a_i$ are in the $(N,N_f)$ representation of
$\mathop{\rm SU}(N)\times \mathop{\rm SU}(N_f)$, $\bar q^a_i$ in the
$(\bar N, \bar N_f)$ and $M^i_j$ in the adjoint representation of
$\mathop{\rm SU}(N_f)$.  In a string context, the quark fields come
from open strings stretching between $N$ D3 and $N_f$ D7-branes.  The
meson fields $M$ are interpreted as the position of the D7-branes.
The D3-branes are placed at singular points in an orientifold
compactification which breaks supersymmetry down to N=1. All the
modular weights of the quarks and mesons should vanish, to prevent the
appearance of a no-scale instability of the vacuum~\cite{pok,scru}. We
neglect the moduli fields corresponding to the motion of the D3-branes
off the singular points and the r\^ole of fractional branes. Although
these moduli could destabilise the vacuum, the study of their dynamics
is left for future work.  We also consider that all the branes are
placed in the bulk of the compactification.  Of course, explicit
constructions should explain the form of the superpotential, e.g.
trilinear couplings appear naturally in quiver models, see
\cite{oo,uf,ka} for more detail.

The K\"ahler potential follows from the results in
section~\ref{s:Kahler}.  For the D7 meson, the modular weight is 0.
For the field joining the D3 and the D7, the brane intersection has
codimension 8, with 6 directions orthogonal to the D3 and 2 to the D7. Now
as the D3 is glued at a fixed point, it is not free to move and
therefore there are effectively only 2 degrees of motion as for a
D7-brane. Hence the modular weight is 0 too.  Expanding in small
fields compared to the Planck scale, this gives an additive
contribution to the K\"ahler potential as already obtained in~\cite{pok}
\begin{equation}
K= \vert q\vert^2 + \vert \tilde q \vert^2 + \vert M\vert^2 \, .
\end{equation}
This justifies the choice of meson fields as the position of D7
branes and the open strings between the D3 and D7-branes as quark
fields. If their modular weights were not zero, the vacuum of the
theory would have been marginally unstable \cite{pok,scru}.

Following the strategy in subsection~\ref{s:W}, we factor out an
overall $\e^{4A_0}$ from the K\"ahler potential, perform a
K\"ahler transformation, and rescale all the brane fields $\phi_i
\to \e^{2A_0} \phi_i$.  This results in a K\"ahler potential from
which all warp factors have been removed; in particular, the ISS
fields $q,\bar q$ and $M$ have canonical kinetic terms.  All
warping effects reside in the superpotential, which for the ISS
sector becomes
\be
W_{\rm ISS} = h \, {\rm tr} (\bar q M q) - h \mu^2 {\rm tr} (M) \, ,
\ee
with $h = \tilde{h}$ not picking up any extra warping dependence,
and $\mu^2 = \e^{-4A_0} \tilde \mu^2$.  At the global supersymmetry
level, the ISS sector has a supersymmetry breaking vacuum
\begin{equation}
M=0 \, , \qquad q^a_i= \tilde q^i_a= \mu \delta^i_a \, .
\end{equation}
The vacuum energy is
\begin{equation}
V_{\rm ISS}  = (N-N_f) h^2 \mu^4 \, .
\end{equation}
Let us now couple the ISS sector to the rest of the model including
the consistent $D$-term. As $\mu\ll \mpl$ and expanding in the small
VEVs of the ISS sector, the leading terms in the scalar potential are
\begin{equation}
V= \frac{1}{(T+\bar T)^3}
\bigg[V(q,\tilde q,M) + V(T, \zeta) \bigg]
\end{equation}
where we have taken into account that $Y\approx 1$ when the VEV of
$\zeta$ is small compared to the Planck scale.  The potential
$V(q,\tilde q,M)$ is the global supersymmetry potential in the ISS
sector. In the large $T$ expansion, the minimum of this potential is
obtained by imposing a separate minimum for $V(T,\chi^i)$ and
$V_{\rm ISS}$ (the corrections terms to the $T$ equation coming from the
ISS sector are in $1/T^4$).  As a result, the cosmological constant at
the minimum is given by
\begin{equation}
V_{\rm min} \approx \frac{(N-N_f) h^2 \mu^4 -3 W_0^2}{8T^3}
\end{equation}
where we have neglected the $\zeta_0$ contribution.  The minimum can
be set to zero provided
\begin{equation}
h\mu^2 \approx \sqrt \frac{3}{N-N_f} \vert W_0 \vert \, .
\label{minISS}
\end{equation}

Let us restore the warp factors. The Minkowski minimum is obtained by
balancing $h \mu^2$ with $W_0$ in \eref{minISS} above.  Since $W_0
\propto \e^{-3A_0}$ this can be achieved only if $h\tilde \mu^2$
scales as $\e^{A_0}$. If this is not the case, the gravitino mass
cannot be warped down. However, if it is the case, the gravitino mass
$m_{3/2} \approx \vert W_0\vert/(2 T_0)^{3/2}$ can be in the TeV range
if $h\mu^2 \sim W_0 \sim 10^{-15}$. Using the warping of $W_0 \propto
\e^{-3A_0}$, this translates into a constraint on $A_0$:
\begin{equation}
\e^{-A_0}\approx 10^{-5} \, .
\end{equation}
We have thus found that a relatively small warping is enough to
lead to a small gravitino mass once a consistent $D$-term model is
coupled to an ISS sector. Moreover, this result depends on the
warping dependence of the mass term of the ISS uplifting sector
and would allow to discriminate explicit ISS uplifting procedures
leading to LHC testable supersymmetry breaking schemes.

\section{Conclusions}

In this paper we analysed the low energy supergravity action in warped
spacetimes with matter on D3 and (intersecting) D7-branes. We
dimensionally reduced the action to find the warping dependence of the
various terms in the 4D action. A central assumption in our setup is
that the throat dominates the volume of the 6D compactified space.
This is valid if the volume modulus $T$, which parameterises the bulk
volume $V_{\rm bulk}^{2/3} \sim T$, is stabilised at moderately large
volumes: $T \ll \e^{4 A_0}$, with $A_0$ the warp factor in the tip of
the throat.

The 4D action for matter on bulk D3 and D7-branes (wrapped around
4-cycles which lie entirely in the bulk) is warping independent.  The
warping only enters via the modulus field $T$, which is a truly 
ten-dimensional field, and thus feels the total 6D volume of compactified
space. This is translated into a scaling
down of the $F$-term potential $V_F \propto \e^{-6A_0}$ via a K\"ahler
transformation.  In contrast, the $D$-term is warping independent.

To study the effect of warping on moduli stabilisation and
supersymmetry breaking we applied it to simple string inspired models
with or without $\alpha'$ corrections in the K\" ahler
potential~\cite{dm}. In these models the moduli sector arises from
matter on magnetised bulk D7-branes, and has a symmetry group
$\mathop{\rm SU}(N) \times \mathop{\rm {}U}(1)$. The chiral matter
content is a meson field and an $\mathop{\rm SU}(N)$ singlet;  the
volume modulus $T$ is charged under the $\mathop{\rm {}U}(1)$. Strong
$\mathop{\rm SU}(N)$ gauge dynamics gives a non-perturbative potential
that stabilises the volume modulus at $\Or(1)$ values. The minimum is
uplifted by $F$- and $D$-terms.

The supersymmetric standard model resides on intersecting D7-branes.
Gauge couplings of order one dictates that these branes are
located in the bulk. Supersymmetry breaking by the moduli sector is
transferred to the standard model fields by gravitational
interactions. Low scale SUSY breaking with a TeV scale gravitino mass
requires both the $F$- and $D$-term to be small compared to the Planck
scale. Since the $D$-term is warping independent, this is only possible
in models in which the $D$-term (nearly) cancels $D \approx 0$.

In the presented models,  it is indeed possible to find $D \approx 0$ 
solutions. In general the resulting minimum is anti de Sitter,
although for some parameter choices non-zero $F$-terms are enough to
raise the minimum to a Minkowski vacuum. Low scale SUSY cannot be
achieved  by turning on a moderate warping. By coupling to an ISS
sector we find that low energy SUSY can be achieved. The AdS vacua
is uplifted via the ISS sector for moderate warping. This results
in a small gravitino mass for warping of the order $10^{-5}$.

\acknowledgments
ACD, SCD thank CEA Saclay for their hospitality.  ACD and MP are also
grateful to the Galileo Galilei Institute and INFN for hospitality and
partial support. For financial support, SCD and RJ thank the
Netherlands Organisation for Scientific Research (NWO), MP thanks FOM,
ACD thanks PPARC for partial support, and PhB acknowledges support
from RTN European programme MRN-CT-2004-503369.

\appendix

\section{Uplifting and stabilisation with light quarks}
\label{a:stephen}

In this Appendix we give the details of the expansion performed in
subsection~\ref{s:exp}.
Expanding the potential around the zeroth order solution
\eref{chiral0} to first order in $\eps$ \eref{expansion} gives
\be V_D = \frac{9 \gs^2}{32 k_X T_0^3}
\left[\epsilon^2 D_1^2 + \Or(\epsilon^3)\right] \, , \qquad V_1 =
\frac{3 b m^2 \zeta_0^{2q}}{8 a^2 T_0^5} \epsilon^3 F_1^2 + \cdots \, , 
\ee 
\bea V_0 &=& V^{(1)}_{\rm min} +\frac{3m \epsilon^2\zeta_0^q}{8
(2+b) a^2 T_0^5} \Bigl\{ m \epsilon \zeta_0^q\left[a \gs D_1 - 2 b F_1
-(2+b+4a + b q)T_1 \right] \nonumber \\ && \hspace{0.5in} {} +a T_0
W_0\left[a \gs D_1 - 2 b F_1 -(8+4b+4a + b q)T_1 \right]\Bigr\} +
\cdots
\label{AV0}
\eea
with
\be
D_1 = 2 \zeta_1 + T_1 \, , \qquad
F_1 = (2+b) \chi_1 + a T_1 +q \zeta_1 \, ,
\ee
where we used gauge invariance $b q = a \gs$.  The $(\cdots)$ in
the above expressions correspond to $\Or(m^2 \zeta_0^{2q}T_0^{-5}
\epsilon^4, m \zeta_0^q T_0^{-4} W_0\epsilon^3, W_0^2 T_0^{-3}
\epsilon^2)$ terms.  As discussed in the main text, the above
potential does not stabilise $T_1$.  The exception to this if
\eref{mincond} is satisfied, leaving open the possibility that $T_1$
could be stabilised by higher order terms the potential.  This is only
possible if $W_0 \sim m \epsilon \zeta_0^q/T_0$, i.e.\ if $W_0 \ll
m$. From subsection~\ref{ssec:W}, we expect $W_0/m \sim
\e^{3A_0+2(2+q)(A_b-A_0)}$, which is indeed tiny when $A_0$ is large
and $A_b$ is close to zero. Without warping the above conditions will
be difficult to fulfil.

The combination $F_1$ is stabilised by the above $F$-terms,
and minimising $V_0+V_1$ with respect to it gives
\be
F_1 = \frac{1}{2+b}\( 1 + \frac{a T_0 W_0}{m \epsilon \zeta_0^q}\) \, .
\ee
The combination $D_1$ is stabilised by $V_0+V_D$ at
\be
D_1 = -\frac{2}{3(2+b)}
\(1 + \frac{a T_0 W_0}{m \epsilon \zeta_0^q}\)
\frac{ k_X m^2 \zeta_0^{2q}\epsilon}{\gs a T_0^2}
\ee
which, assuming~\eref{mincond}, is negligible unless
$W_0^2 \sim \epsilon \gs/k_X$. From now on we will assume, for simplicity,
that $V_D$ dominates the other parts of the potential, and so $D_1=0$.

Setting $V^{(1)}_{\rm min} =0$ to obtain a Minkowski vacuum requires
\eref{cond}, \eref{aT1}.  The condition~\eref{expansion} implies
\be
\frac{1}{(a T_0)^{(2+q)/2}} \gg
\frac{(-2 W_0)}{m}\left(\frac{4}{3bq}\right)^{q/2} \gg
\frac{1}{(a T_0)^{(3+q)/2}} \, .
\label{exp2}
\ee
From~\eref{W3} we find $A/m \sim \e^{2(b+2+q)(A_b-A_0)}$. Combining
this with $\gs \sim \e^{-4 A_b}$ we find
\be a T_0 \sim \(a -\frac{3b q}{4} \) A_0 + (4a-b q)(A_b-A_0) \, .
\ee
For $A_b=0$, $W_0/m \sim e^{-(1+2q) A_0}$ and $a T_0 \sim (bq
-12a)A_0/4$. To satisfy \eref{exp2} and the other consistency
conditions, we need to take $N$ and $N_f$ large. For example, consider
$N=1024$, $N_f=1023$, $k_N= 1/\pi^2$. Then determine $a, b, q, \gs$
from \eref{Aab}, \eref{anom1} and \eref{cond}; this gives $a=8,
b=2046, q=1.986, \gs = 508$. Taking $A_0 =2$, we then obtain $a T_0
\approx 1936$ and $T_0 \approx 242$ from~\eref{aT1}.  The expansion
parameter in this case $\eps = 0.12$, which marginally satisfies
\eref{expansion}.  However, taking $N \sim 10^3$ D7-branes is not a
particularly realistic situation.

For completeness, we also give the higher order expansion of the
potential. To next order in $\epsilon$, we find
\be
V_D = \frac{9 \gs^2}{32k_X T_0^3}
\left[\epsilon^4 D_2^2+\Or(\epsilon^5)\right]
\ee
\bea
V_F = \mathrm{const.} + \frac{3W_0^2\epsilon^2}{2 b (2+b) T_0^3}\biggl\{
&& \hspace{-0.1in}
\left[4(a-1)^2 + (a-2)(2 a-1) b\right] T_1^2
\nonumber \\ && {}
+ (2+b-2a)b\left[\frac{a T_1}{4(2+b)} + D_2\right]
 + \Or(\epsilon) \biggr\}
\label{vf2}
\eea
with $D_2 = 2 \zeta_2 +T_2-3 T_1^2/4$.  The $D$-term part of the
potential stabilises $D_2$. The above solution will only be a minimum
if the coefficient of $T_1^2$ in~\eref{vf2} is positive, otherwise
$T_1$ will not be stabilised. We see that for the above parameters, we
do indeed have a minimum.

\section{Large volume stabilisation with higher order $\alpha'$ corrections}
\label{a:alpha}

After integrating out the quark fields, the effective superpotential
\eref{Wads} has a SUSY AdS min with $F=D=0$.  This conclusion can be
avoided if higher order $\alpha'$ corrections to the K\"ahler
potential play a r\^ole~\cite{quevedoT}.  This allows for the
possibility of large volume stabilisations with $\Re T \gg 1$.  In
this limit the non-perturbative effects are sub-dominant, and the SUSY
preserving condition $F=0$ cannot be satisfied making a Minkowski
minimum is possible.

The $\alpha'$ corrections come from terms of the form
\be
\epsilon^{ABM_1M_2\dots M_8} \epsilon_{AB N_1 N_2\dots n_8 }
R_{M_1M_2}\ ^{N_1N_2}  R_{M_3M_4}\ ^{N_3N_4} R_{M_5M_6}\ ^{N_5N_6}
R_{M_7M_8}\ ^{N_7N_8}
\ee
in the 10D action~\cite{becker}. Upon
dimensional reduction it leads to a correction of the K\"ahler
potential, which becomes (after performing the K\"ahler transformation
\eref{Ktrafo})~\cite{quevedoT,becker}
\be K= -\frac{2}{\kappa_4^2} \ln \Big[ (T +\bar T)^{3/2} + \xi
\Big] -\frac{3}{\kappa_4^2} \ln \Big[ 1-\frac{\kappa_4^2}{3}
|\zeta|^2  \Big] \ee
with
\be \xi = \e^{-4A_0} \tilde \xi \label{xi} \ee
parameterising the $\alpha'$ corrections.  We see that the higher
order curvature term is warped down with respect to the Einstein term,
as was shown in subsection~\ref{ss:sugra}. In the limit $A_0 \to 0$
our results agree with the literature~\cite{quevedo}, and in the
limit $\xi \to 0$ it reduces to our previous result~\eref{K3}.  The
size of the corrections diminishes rapidly for large warping. It is
thus expected that it cannot play a r\^ole in this limit.  As we
will show now, this is indeed the case.

To analyse the system we make a $1/T_R$ expansion with
$T_R = {\rm Re}\, T$.  At lowest order the $F$ and $D$-term potential are
\bea V_F &=& \frac{W_0^2}{(2T_R)^3 Y^3} \left\{ |\zeta|^2 + \frac{3\xi}{2
(2T_R)^{3/2}} + \Or \(\zeta^2, \frac{\xi}{T_R^{3/2}}\)^2
 \right\} + V_{\rm non-pert}
\nonumber\\
V_D &=& \frac{9 \gs^2}{32 k_X T_R^3} \(1-\frac{4T_R}{3\gs Y}
|\zeta|^2 \)^2 \eea
with $Y = (1 - |\zeta|^2/3)$ and
$V_{\rm non-pert}= V_F- e^K(K_{\bar \imath} K^{\bar \imath j} K_j -3)W_0^2$,
i.e.\ $V_{\rm non-pert}$ includes all terms with
non-perturbative $\e^{-(2a T)/(b+2)}$ factors.  Both terms come in at the 
same order, which gives a $D \neq 0$ minimum, with high scale SUSY
breaking.  This can be avoided if $W_0^2 \ll 1/{T_R}$.  Then the
$D$-term potential dominates, and $\zeta$ will adjust itself to cancel
it:
\be
|\zeta_0|^2 = \frac{3 \gs}{4T_R} \, .
\ee
One linear combination of the phase fields is fixed by the
non-perturbative potential, the orthogonal contribution remains massless
and gets eaten by the anomalous $\mathop{\rm {}U}(1)$.  $T_R$ is fixed
at higher order.
Write $\zeta =\zeta_0 [1 + \zeta_1/T_R + \Or(1/T_R^{2})]$,
and expand the potential to lowest order
\bea V_F &=& \frac{W_0^2}{(2T_R)^3} \left[ \frac{3 \gs}{4T_R}
\(1+ \frac2{T_R} \zeta_1 \) +\frac{3\xi}{2 (2T_R)^{3/2}}
\right] + V_{\rm non-pert} + \cdots
\nonumber\\
V_D &=& \frac{9 \gs^2}{8 k_X T_R^5} \(\frac{\gs}{8}+ \zeta_1\)^2 \, .
\eea
The field $|\zeta|$ is fixed at this and successively higher orders by
the condition $D=0$.  Although $|\zeta|$ also appears in the $F$-term
potential, it will always be at higher order.  The volume modulus is
stabilised by the $F$-term potential, which requires $\xi < 0$.  The
$F$-term potential is minimised at
\be
T_R = \frac{81 \xi^2}{128 \, \gs^2}
\ee
at a value
\be V_{\rm min}  =  \frac{\gs W_0^2}{96 T_R^4} \, .
\ee
The minimum is dS, not a viable cosmological solution.  The warp
dependence of $\xi$~\eref{xi} does not help here, as $T_R$ will be
small for large warping, and our expansion in $1/T_R$ will break down.
The difference between our results and those of~\cite{quevedoT}, who
did find a AdS/Minkowski minimum, is the different modular weight for
$\zeta$.

\end{document}